\documentclass{article}
\usepackage{amsmath}
\usepackage{graphicx}
\usepackage{amsfonts}
\usepackage{amssymb}
\input epsf
\advance \topmargin by -\headheight
\advance \topmargin by -\headsep
\evensidemargin \oddsidemargin
\begin{document}

\title{Introduction to stochastic error correction methods}
\author{Dean Lee$^{a}\thanks{dlee@physics.umass.edu}$, Nathan Salwen$^{b}%
\thanks{salwen@physics.harvard.edu}$, Mark Windoloski$^{a}%
\thanks{markw@het2.physics.umass.edu}$\\$^{a}$Dept. of Physics, Univ. of Massachusetts, Amherst, MA 01003\\$^{b}$Dept. of Physics, Harvard Univ., Cambridge, MA 02138\\\\}
\maketitle
\begin{abstract}
We propose a method for eliminating the truncation error associated with any
subspace diagonalization calculation. \ The new method, called stochastic
error correction, uses Monte Carlo sampling to compute the contribution of the
remaining basis vectors not included in the initial diagonalization. \ The
method is part of a new approach to computational quantum physics which
combines both diagonalization and Monte Carlo techniques. \ 
\end{abstract}

\section{Introduction}

In \cite{qse} a new approach was proposed for finding the low-energy
eigenstates of very large or infinite-dimensional quantum Hamiltonians. \ This
proposal combines both diagonalization and Monte Carlo methods, each being
used to solve a portion of the problem for which the technique is most
efficient. \ The first part of the proposal is to diagonalize the Hamiltonian
restricted to a subspace containing the most important basis vectors for each
low energy eigenstate. \ This may be accomplished either through variational
techniques or an \textit{ab initio} method such as quasi-sparse eigenvector
(QSE) diagonalization.\ \ The second step is to include the contribution of
the remaining basis vectors\ by Monte Carlo sampling. \ The use of
diagonalization allows one to consider systems with fermion sign oscillations
and extract information about wavefunctions and excited states. \ The use of
Monte Carlo provides tools to handle the exponential increase in the number of
basis states for large volume systems.

The first half of this proposal was discussed in \cite{qse}. \ An adaptive
diagonalization algorithm known as the quasi-sparse eigenvector method was
introduced to find the most important basis vectors for each low energy
eigenstate. \ This technique is especially valuable when little is known about
the low energy states. \ It is also the only method available which can handle
non-orthogonal bases, non-Hermitian Hamiltonians, infinite dimensional
systems, and which can find several low energy states with like quantum
numbers simultaneously. \ In this paper we discuss the second half of the
diagonalization/Monte Carlo scheme. \ We introduce several new Monte Carlo
techniques which we call stochastic error correction (SEC). \ There are two
general varieties of stochastic error correction, methods based on a series
expansion and those which are not. \ The series method starts with an
eigenvector of the Hamiltonian restricted to some starting subspace and then
includes the contribution of the remaining basis states as terms in an ordered
expansion. \ The idea is to form a perturbative expansion centered around a
good non-perturbative starting point.

As an example of a non-series method we discuss a technique called the
stochastic Lanczos method. \ This method again starts with eigenvectors of a
Hamiltonian submatrix. \ Using these as starting vectors, we define Krylov
vectors, $\left|  j\right\rangle ,H\left|  j\right\rangle ,H^{2}\left|
j\right\rangle \cdots$, similar to standard Lanczos diagonalization. \ The new
ingredient is that matrix elements between Krylov vectors, $\left\langle
j^{\prime}\right|  H^{n}\left|  j\right\rangle ,$ are computed using matrix
diffusion Monte Carlo. \ Since the method does not rely on a series expansion,
it has the advantage that the starting vectors need not be close to the exact eigenvectors.

One can generate a large class of stochastic error correction methods based on
other non-series algorithms, various ways of resumming the series expansion,
or combinations of the two techniques. \ In this introductory paper we
concentrate on describing the basic principles and implementation of the
series and non-series approaches. \ We also present three test problems which
demonstrate the potential of the new approach for a range of different
problems. \ In the first example we determine the low energy spectrum of
$\phi_{2+1}^{4}$ theory using QSE diagonalization and first order corrections
using the series method. \ In the second example we find the low energy
spectrum of compact $U(1)$ in $2+1$ dimensions using the stochastic Lanczos
method. \ In the last example we find the ground state of the $2+1$
dimensional Hubbard model using QSE diagonalization and first order series
stochastic error correction. \ In each case we compare with published results
in the literature. \ We conclude with a summary and some general comments on
the new computational scheme.

\section{Series method}

Let $\left|  i\right\rangle $ be the eigenvectors of a Hamiltonian $H$
restricted to some subspace $S$. \ Let $\left|  A_{j}\right\rangle $ be the
remaining basis vectors in the full space not contained in $S$. \ We can
represent $H$ as
\begin{equation}
\left[
\begin{array}
[c]{cccccc}%
\lambda_{1} & 0 & \cdots & \left\langle 1\right|  H\left|  A_{1}\right\rangle
& \left\langle 1\right|  H\left|  A_{2}\right\rangle  & \cdots\\
0 & \lambda_{2} & \cdots & \left\langle 2\right|  H\left|  A_{1}\right\rangle
& \left\langle 2\right|  H\left|  A_{2}\right\rangle  & \cdots\\
\vdots & \vdots & \ddots & \vdots & \vdots & \cdots\\
\left\langle A_{1}\right|  H\left|  1\right\rangle  & \left\langle
A_{1}\right|  H\left|  2\right\rangle  & \cdots &  E\cdot\lambda_{A_{1}} &
\left\langle A_{1}\right|  H\left|  A_{2}\right\rangle  & \cdots\\
\left\langle A_{2}\right|  H\left|  1\right\rangle  & \left\langle
A_{2}\right|  H\left|  2\right\rangle  & \cdots & \left\langle A_{2}\right|
H\left|  A_{1}\right\rangle  & E\cdot\lambda_{A_{2}} & \cdots\\
\vdots & \vdots & \vdots & \vdots & \vdots & \ddots
\end{array}
\right]  .
\end{equation}
We have used Dirac's bra-ket notation to represent the terms of the matrix.
\ In cases where the basis is non-orthogonal or the Hamiltonian is
non-Hermitian, the precise meaning of terms such as $\left\langle
A_{1}\right|  H\left|  1\right\rangle $ is the action of the dual vector to
$\left|  A_{1}\right\rangle $ upon the vector $H\left|  1\right\rangle $. \ We
have written the diagonal terms for the basis vectors $\left|  A_{j}%
\right\rangle $ with an explicit factor $E$ for reasons to be explained shortly.

Let us assume that $\left|  1\right\rangle $ is close to some exact
eigenvector of $H$ which we denote as $\left|  1_{\text{full}}\right\rangle $.
\ More concretely we assume that the components of $\left|  1_{\text{full}%
}\right\rangle $ outside $S$ are small enough so that we can expand in inverse
powers of the introduced parameter $E.$ \ In order to simply the expansion we
choose to shift the diagonal entries so that $\lambda_{1}=0.$

The series method of stochastic error correction is based on the $E^{-1}$
expansion,
\begin{equation}
\left|  1_{\text{full}}\right\rangle \propto\left[
\begin{array}
[c]{c}%
1\\
c_{2}^{\prime}E^{-1}+c_{2}^{\prime\prime}E^{-2}+\cdots\\
\vdots\\
c_{A_{1}}^{\prime}E^{-1}+c_{A_{1}}^{\prime\prime}E^{-2}+\cdots\\
c_{A_{2}}^{\prime}E^{-1}+c_{A_{2}}^{\prime\prime}E^{-2}+\cdots\\
\vdots
\end{array}
\right]  , \label{unnorm}%
\end{equation}%
\begin{equation}
\lambda_{\text{full}}=\lambda_{1}^{\prime}E^{-1}+\lambda_{1}^{\prime\prime
}E^{-2}\cdots.
\end{equation}
It is convenient to choose the normalization of the eigenvector such that the
$\left|  1\right\rangle $ component remains 1. \ The convergence of the
expansion is controlled by the proximity of $\left|  1\right\rangle $ to
$\left|  1_{\text{full}}\right\rangle $. \ If $\left|  1\right\rangle $ is not
at all close to $\left|  1_{\text{full}}\right\rangle $ then it will be
necessary to use a non-series method such as the stochastic Lanczos method
discussed in the next section.

At first order in $E^{-1}$ we find
\begin{equation}
c_{A_{j}}^{\prime}=-\tfrac{\left\langle A_{j}\right|  H\left|  1\right\rangle
}{\lambda_{A_{j}}}%
\end{equation}%
\begin{equation}
\lambda_{1}^{\prime}=-\sum_{j}\tfrac{\left\langle 1\right|  H\left|
A_{j}\right\rangle \left\langle A_{j}\right|  H\left|  1\right\rangle
}{\lambda_{A_{j}}} \label{firs}%
\end{equation}%
\begin{equation}
c_{j}^{\prime}=\tfrac{1}{\lambda_{j}}\sum_{k}\tfrac{\left\langle j\right|
H\left|  A_{k}\right\rangle \left\langle A_{k}\right|  H\left|  1\right\rangle
}{\lambda_{A_{k}}}.
\end{equation}
At second order the contributions are
\begin{equation}
c_{A_{j}}^{\prime\prime}=\sum_{k\neq j}\tfrac{\left\langle A_{j}\right|
H\left|  A_{k}\right\rangle \left\langle A_{k}\right|  H\left|  1\right\rangle
}{\lambda_{A_{j}}\lambda_{A_{k}}}-\sum_{l\neq1}\sum_{k}\tfrac{\left\langle
A_{j}\right|  H\left|  l\right\rangle \left\langle l\right|  H\left|
A_{k}\right\rangle \left\langle A_{k}\right|  H\left|  1\right\rangle
}{\lambda_{A_{j}}\lambda_{l}\lambda_{A_{k}}}%
\end{equation}
\qquad%
\begin{equation}
\lambda_{1}^{\prime\prime}=\sum_{j}\sum_{k\neq j}\tfrac{\left\langle 1\right|
H\left|  A_{j}\right\rangle \left\langle A_{j}\right|  H\left|  A_{k}%
\right\rangle \left\langle A_{k}\right|  H\left|  1\right\rangle }%
{\lambda_{A_{j}}\lambda_{A_{k}}}-\sum_{j}\sum_{l\neq1}\sum_{k}\tfrac
{\left\langle 1\right|  H\left|  A_{j}\right\rangle \left\langle A_{j}\right|
H\left|  l\right\rangle \left\langle l\right|  H\left|  A_{k}\right\rangle
\left\langle A_{k}\right|  H\left|  1\right\rangle }{\lambda_{A_{j}}%
\lambda_{l}\lambda_{A_{k}}}%
\end{equation}%

\begin{align}
c_{m}^{\prime\prime}  &  =-\sum_{j}\sum_{k\neq j}\tfrac{\left\langle m\right|
H\left|  A_{j}\right\rangle \left\langle A_{j}\right|  H\left|  A_{k}%
\right\rangle \left\langle A_{k}\right|  H\left|  1\right\rangle }{\lambda
_{m}\lambda_{A_{j}}\lambda_{A_{k}}}+\sum_{j}\sum_{l\neq1}\sum_{k}%
\tfrac{\left\langle m\right|  H\left|  A_{j}\right\rangle \left\langle
A_{j}\right|  H\left|  l\right\rangle \left\langle l\right|  H\left|
A_{k}\right\rangle \left\langle A_{k}\right|  H\left|  1\right\rangle
}{\lambda_{m}\lambda_{A_{j}}\lambda_{l}\lambda_{A_{k}}}\\
&  -\tfrac{1}{\lambda_{m}}\left[  \sum_{j}\tfrac{\left\langle 1\right|
H\left|  A_{j}\right\rangle \left\langle A_{j}\right|  H\left|  1\right\rangle
}{\lambda_{A_{j}}}\right]  \tfrac{1}{\lambda_{m}}\left[  \sum_{k}%
\tfrac{\left\langle m\right|  H\left|  A_{k}\right\rangle \left\langle
A_{k}\right|  H\left|  1\right\rangle }{\lambda_{A_{k}}}\right]  .\nonumber
\end{align}
These contributions can be calculated by straightforward Monte Carlo sampling.
\ All that is required is an efficient way of generating random basis vectors
$\left|  A_{k}\right\rangle $ with known probability rates. \ Let
$P(A_{\text{trial}})$ denote the probability of selecting $\left|
A_{\text{trial}}\right\rangle $ on a given trial. \ If for example we are
calculating the first order correction to the eigenvalue, then we have%
\begin{align}
\lambda_{1}^{\prime}  &  =-\sum_{j}\tfrac{\left\langle 1\right|  H\left|
A_{j}\right\rangle \left\langle A_{j}\right|  H\left|  1\right\rangle
}{\lambda_{A_{j}}}\\
&  =-\lim_{N\rightarrow\infty}\tfrac{1}{N}\sum_{i=1,\cdots,N}\tfrac
{\left\langle 1\right|  H\left|  A_{\text{trial}(i)}\right\rangle \left\langle
A_{\text{trial}(i)}\right|  H\left|  1\right\rangle }{\lambda_{A_{\text{trial}%
(i)}}P(A_{\text{trial}(i)})}.\nonumber
\end{align}

\section{Stochastic Lanczos}

We now consider a method called stochastic Lanczos which does not require the
starting vectors to be close to exact eigenvectors of $H.$ \ This is essential
if the eigenvectors of $H$ are not quasi-sparse and require extremely large
numbers of basis states to represent accurately.

Let $V$ be the full Hilbert space for our system. \ As in the previous section
let $S$ be the subspace over which we have diagonalized $H$ exactly. \ Let
$P_{S}$ be the projection operator for $S$ and let $\lambda_{j}$ and $\left|
j\right\rangle $ be the eigenvalues and eigenvectors of $H$ restricted to $S$
so that
\begin{equation}
P_{S}HP_{S}\left|  j\right\rangle =\lambda_{j}\left|  j\right\rangle .
\end{equation}
\ Let $Z$ be an auxiliary subspace, one which contains $S$ but excludes very
high-energy states. Let $P_{Z}$ be the projection operator for $Z$. \ We will
choose $Z$ such that $P_{Z}HP_{Z}$ is bounded above. \ Let$\ a$ be a real
constant which is greater than the midpoint of the minimum and maximum
eigenvalues of $P_{Z}HP_{Z}$. \ As $n\rightarrow\infty$ the operator $\left[
P_{Z}(H-a)P_{Z}\right]  ^{n}$ maps any given state in $Z$ to the corresponding
lowest-energy eigenvector of $P_{Z}HP_{Z}$\ with non-zero overlap.

The stochastic Lanczos method uses the operators $\left[  P_{Z}(H-a)P_{Z}%
\right]  ^{n}$ to approximate the low-energy eigenvalues and eigenvectors of
$P_{Z}HP_{Z}$. \ The goal is to diagonalize $H$ in a subspace spanned by vectors%

\begin{equation}
\left|  d,j\right\rangle =\left[  P_{Z}(H-a)P_{Z}\right]  ^{d}\left|
j\right\rangle , \label{slbasis}%
\end{equation}
for several values of $d$ and $j$. \ This requires calculating $\left\langle
d^{\prime},j^{\prime}\right|  \left.  d,j\right\rangle $ and $\left\langle
d^{\prime},j^{\prime}\right|  H\left|  d,j\right\rangle $. \ If our
Hamiltonian matrix is Hermitian, both of these terms can be written in the
general form
\begin{equation}
\left\langle j^{\prime}\right|  \left[  P_{Z}(H-a)P_{Z}\right]  ^{n}\left|
j\right\rangle .
\end{equation}
Therefore it suffices to determine the matrix
\begin{equation}
A_{n}\equiv P_{S}\left[  P_{Z}(H-a)P_{Z}\right]  ^{n}P_{S}.
\end{equation}
For non-orthogonal bases and non-Hermitian Hamiltonians, the only change is
that we use vectors
\begin{equation}
\left[  P_{Z}(H-a)P_{Z}\right]  ^{d}\left|  j\right\rangle
\end{equation}
to generate approximate right eigenvectors of $H$ and vectors in the dual
space
\begin{equation}
\left\langle j\right|  \left[  P_{Z}(H-a)P_{Z}\right]  ^{d}%
\end{equation}
to produce approximate left eigenvectors. \ Adding and subtracting
$P_{S}(H-a)P_{S}$, we can rewrite
\begin{equation}
A_{n}=P_{S}\left[  P_{Z}(H-a)P_{Z}-P_{S}(H-a)P_{S}+P_{S}(H-a)P_{S}\right]
^{n}P_{S}.
\end{equation}
$A_{n}$ can now be evaluated recursively as
\begin{equation}
A_{n+1}=B_{n+1}+\sum_{m=0,\cdots,n}B_{m}(H-a)A_{n-m},
\end{equation}
where
\begin{equation}
B_{n}=P_{S}\left[  P_{Z}(H-a)P_{Z}-P_{S}(H-a)P_{S}\right]  ^{n}P_{S}.
\end{equation}

The components of $B_{n}$ are computed by matrix diffusion Monte Carlo. \ One
could also directly evaluate the components of $A_{n}$. \ However the
calculation for $B_{n}$ eliminates the need to sample the matrix
$P_{S}(H-a)P_{S}$, which is already known. \ Any general matrix product
$M^{(1)}M^{(2)}\cdots M^{(n)}$ is a sum of degree $n$ monomials,
\begin{equation}
\left[  M^{(1)}M^{(2)}\cdots M^{(n)}\right]  _{jk}=\sum_{i_{1},\cdots i_{n-1}%
}M_{ji_{1}}^{(1)}M_{i_{1}i_{2}}^{(2)}\cdots M_{i_{n-1}k}^{(n)}. \label{sum}%
\end{equation}
We can interpret (\ref{sum}) as a sum over paths through the set of basis
vectors of $Z,$%
\begin{equation}
\left|  j\right\rangle \rightarrow\left|  i_{1}\right\rangle \rightarrow
\cdots\rightarrow\left|  i_{n-1}\right\rangle \rightarrow\left|
k\right\rangle \text{,}%
\end{equation}
with an associated weight $M_{ji_{1}}^{(1)}M_{i_{1}i_{2}}^{(2)}\cdots
M_{i_{n-1}k}^{(n)}$. \ The components of $B_{n}$ are sampled using ensembles
of random walkers. \ We refer the interested reader to \cite{kosztin} for a
review of methods in diffusion Monte Carlo.

We end the section with a discussion of the fermion sign problem. \ The sign
problem is a general issue for any Monte Carlo calculation. \ For a system
with sign oscillations the evaluation of a Euclidean-time Green's function
involves sums
\begin{equation}
\sum_{i}x_{i}%
\end{equation}
with the property that
\begin{equation}
\frac{\sum_{i}x_{i}}{\sum_{i}\left|  x_{i}\right|  }\sim\exp(-c\cdot V\cdot
T),
\end{equation}
where $V$ is the volume, $T$ is the Euclidean time, and $c$ is a positive
constant. \ We will refer to this term as the cancellation ratio. \ The
exponential dependence on $V$ and $T$ makes computations difficult even for
small systems.

The sign problem will affect the calculation of $B_{n}$ in the stochastic
Lanczos algorithm and terms in the series method discussed in the previous
section. \ The effect however is different from the sign problem in typical
Monte Carlo Green's function calculations. \ Stochastic error correction is a
calculation of eigenvalues and eigenvectors rather than a sampling of the
partition function or the time evolution of a given initial state. \ Therefore
the quantity of interest is not $\exp(-HT)$ but the action of $H$ or $H^{n}$
on approximate eigenvectors of $H$. \ Due to homogeneity in $H$ the explicit
volume dependence does not appear in the cancellation ratio. \ Instead we
find
\begin{equation}
\frac{\sum_{i}x_{i}}{\sum_{i}\left|  x_{i}\right|  }\sim\exp(-k\cdot n),
\end{equation}
where $k$ is a positive constant. \ The sign problem will return if the
starting point of the SEC calculation is very poor and it becomes necessary to
use $n$ such that $k\cdot n$ is large. \ However in many cases $k\cdot n$ can
be kept small even for large $n$ since the most important part of the
Hamiltonian, $P_{S}HP_{S},$ is diagonalized exactly. \ In short the sign
problem is less severe because stochastic error correction uses the result of
subspace diagonalization as its starting point.

\section{$\phi^{4}$ theory in $2+1$ dimensions}

The first example we consider is $\phi^{4}$ theory in $2+1$ dimensions near
the $\phi\rightarrow-\phi$ symmetry restoration phase transition. \ We will
use QSE diagonalization and the series version of stochastic error correction
to probe the low energy spectrum of the theory on both sides of the phase transition.

In \cite{mag} Magruder demonstrated the existence of a phase transition in
$\phi_{2+1}^{4}$ by extending Chang's duality argument for $\phi_{1+1}^{4}$.
\ The statement of the main result is as follows. \ Consider the two Lagrange
densities%
\begin{align}
\mathcal{L}_{+}  &  =\tfrac{1}{2}\partial_{\nu}\phi\partial^{\nu}\phi
-\tfrac{1}{2}\mu_{+}^{2}\phi^{2}-\tfrac{g}{4!}\phi^{4}+\tfrac{1}{2}\delta
_{\mu_{+}}^{2}\phi^{2}\\
\mathcal{L}_{-}  &  =\tfrac{1}{2}\partial_{\nu}\phi\partial^{\nu}\phi
+\tfrac{1}{4}\mu_{-}^{2}\phi^{2}-\tfrac{g}{4!}\phi^{4}+\tfrac{1}{2}\delta
_{\mu_{-}}^{2}\phi^{2}.
\end{align}
The counterterm $\delta_{\mu_{+}}^{2}$ is defined so that in the
$\mathcal{L}_{+}$ system the $\phi$ self-energy graphs vanish at zero-momentum
up to two-loop order. \ By shifting the field%
\begin{equation}
\phi=\phi^{\prime}+\sqrt{\tfrac{3\mu_{-}^{2}}{g}}%
\end{equation}
we note that the same counterterm $\delta_{\mu_{-}}^{2}$ (same mass dependence
but $\mu_{+}$ replaced by $\mu_{-}$) is also sufficient to renormalize
$\mathcal{L}_{-}$. \ By equating $\mathcal{L}_{+}$ and $\mathcal{L}_{-}$ we
obtain a duality constraint between the two theories. \ One feature of this
constraint is that the $g\rightarrow\infty$ limit of $\mathcal{L}_{-}$ is
mapped to the $g\rightarrow0$ limit of $\mathcal{L}_{+}$. \ Therefore
$\mathcal{L}_{-}$, whose reflection symmetry $\phi\rightarrow-\phi$ is broken
at small $g$, must eventually reach the symmetric phase for sufficiently large
coupling.\footnote{The $g\rightarrow\infty$ limit of $\mathcal{L}_{+}$ is
mapped to the $g\rightarrow\infty$ limit of $\mathcal{L}_{-}$ and so there is
no analogous argument for a phase transition in $\mathcal{L}_{+}$. \ Numerical
calculations indicate that there is no phase transition for $\mathcal{L}_{+}$
\cite{markw}\cite{thesis}.}

The $\mathcal{L}_{-}$ phase transition was studied using quasi-sparse
eigenvector diagonalization with stochastic error correction. \ Quantities
such as the critical coupling, critical exponents, and the low lying energy
spectrum were studied and, where possible, compared with Monte Carlo results.
\ A full discussion methods and results are presented in \cite{thesis}. \ We
will very briefly summarize some of the results below.

The two spatial dimensions of our system are taken to be a periodic box of
size $2L$ by $2L$. We will use the modal field formalism to describe the
Hamiltonian for the theory.\footnote{We refer the reader to\ \cite{modal} for
a short introduction.} \ In the following we let the vectors $\vec{n}$
represent ordered integer pairs $(n_{x},n_{y})$ such that $\left|
n_{x}\right|  ,\left|  n_{y}\right|  \leq N_{\max}$. \ The parameter $N_{\max
}$ corresponds with a momentum cutoff scale of $\Lambda=N_{\max}\pi/L$. \ The
modal field Hamiltonian has the form\footnote{Counterterms were calculated
using finite volume perturbation theory.}%

\begin{align}
H  &  =\sum_{\vec{n}}\left[  -\tfrac{1}{2}\tfrac{\partial}{\partial
q_{-\vec{n}}}\tfrac{\partial}{\partial q_{\vec{n}}}+\tfrac{1}{2}\left(
\tfrac{\vec{n}^{2}\pi^{2}}{L^{2}}-\tfrac{\mu^{2}}{2}\right)  -\tfrac
{6b}{(2L)^{2}}\tfrac{g}{4!}+\tfrac{48}{(2L)^{4}}\left(  \tfrac{g}{4!}\right)
^{2}\alpha_{\vec{n}}\right]  q_{-\vec{n}}q_{\vec{n}}\\
&  +\tfrac{1}{(2L)^{2}}\tfrac{g}{4!}\sum_{\vec{n}_{1}+\vec{n}_{2}+\vec{n}%
_{3}+\vec{n}_{4}=0}q_{\vec{n}_{1}}q_{\vec{n}_{2}}q_{\vec{n}_{3}}q_{\vec{n}%
_{4}}\nonumber
\end{align}
where
\begin{equation}
b=\sum_{\vec{n}}\tfrac{1}{2\omega_{\vec{n}}},\qquad\omega_{\vec{n}}%
=\sqrt{\tfrac{\vec{n}^{2}\pi^{2}}{L^{2}}+\mu^{2}},
\end{equation}
and%
\begin{equation}
\alpha_{\vec{n}}=\sum_{\vec{n}_{1},\vec{n}_{2}}\tfrac{1}{4\omega_{\vec{n}_{1}%
}\omega_{\vec{n}_{2}}\omega_{\vec{n}-\vec{n}_{1}-\vec{n}_{2}}(\omega_{\vec
{n}_{1}}+\omega_{\vec{n}_{2}}+\omega_{\vec{n}-\vec{n}_{1}-\vec{n}_{2}})}.
\end{equation}

In Figure 1 we have plotted the lowest energy eigenstates in the rest frame
for $N_{\max}=10$ and $L=3\pi$ (in units where $\mu=1$). \ This choice of
parameters corresponds with $441$ different momentum modes and a momentum
cutoff scale of $\Lambda=3.33\mu$. \ The states shown in Figure 1 are the
three lowest eigenstates in the even and odd $\phi\rightarrow-\phi$ symmetry
sectors, and the energies are measured relative to the ground state energy.
\ In our calculation QSE diagonalization was used keeping $500$ Fock states,
and the stochastic error correction was computed to first-order using the
series method. \ Error bars shown include statistical error and an estimate of
the contribution from higher order corrections. \ We see clear evidence of a
second-order phase transition near $\frac{g}{4!}=0.9$.\footnote{A more
complete discussion of the critical coupling as well as critical exponents can
be found in \cite{thesis}.} \ We have labelled the energies of the states
according to their physical interpretation in the symmetric phase. \ $E_{1}$
is the energy for the one-particle state, $E_{2}(E_{3})$ is for the
two(three)-particle threshold, and $E_{2}^{\prime}(E_{3}^{\prime})$ is for the
first state above the two(three)-particle threshold. \ At finite volume these
energies are continuous functions of the coupling $g$. \ One feature which was
also observed in $\phi_{1+1}^{4}$ \cite{qse} is the crossing of energy levels
due to the double degeneracy of states in the broken symmetry phase. $\ E_{3}$
is connected to a one-particle state in the broken phase while $E_{2}^{\prime
}$ is connected to a two-particle state. \ The levels $E_{2}^{\prime}$ and
$E_{3}$ therefore cross near the critical point.

Another interesting phenomenon is the appearance of a bound state in the
broken symmetry phase. \ In both the odd and even symmetry sectors we can
measure the ratio of the two-particle to one-particle energies:%
\[
\overset{\text{Table 1}}{%
\begin{tabular}
[c]{|l|l|l|}\hline
$\frac{g}{4!}$ & $E_{2}^{\prime}/E_{2}$ & $E_{3}^{\prime}/E_{3}$\\\hline
$0.2$ & $2.01(4)$ & $1.98(4)$\\\hline
$0.3$ & $2.01(4)$ & $2.05(4)$\\\hline
$0.4$ & $1.95(4)$ & $1.96(4)$\\\hline
$0.5$ & $1.87(4)$ & $1.87(4)$\\\hline
$0.6$ & $1.86(4)$ & $1.82(4)$\\\hline
\end{tabular}
}%
\]
These results are consistent with the binding energies reported in \cite{CHP}
and \cite{CHPZ}, which indicate a ratio of $1.83(3)$ near the critical point.

\section{Compact U(1) in 2+1 dimensions}

Compact U(1) lattice gauge theory in $2+1$ dimensions is a simple but
phenomenologically interesting gauge model. \ It is asymptotically free and in
the usual continuum limit describes massless non-interacting photons. \ On the
other hand if the continuum limit is reached by rescaling the mass gap to
remain constant, one instead finds a confining theory of massive bosons
\cite{gopfert}. \ The Hamiltonian has the form%

\begin{equation}
H=-\sum_{l}\tfrac{\partial^{2}}{\partial A_{l}^{2}}-2x\sum_{P}\cos\theta_{P}.
\label{hamil}%
\end{equation}
In (\ref{hamil}) $A_{l}$ are link gauge fields, $\theta_{P}$ is the sum of the
links circuiting a plaquette,%
\begin{equation}
\theta_{P}=A_{l_{1}}+A_{l_{2}}-A_{l_{3}}-A_{l_{4}},
\end{equation}
and%
\begin{equation}
x=e^{-4}a^{-2}%
\end{equation}
is the strong coupling parameter, which tends to infinity as the lattice
spacing $a$ goes to $0$. \ We follow the notation of \cite{hamer} in which an
overall constant factor of $\frac{e^{2}}{2}$ multiplying the right-hand side
of (\ref{hamil}) is suppressed. \ The energy levels we measure are therefore
in units of $\frac{e^{2}}{2}$.

The diagonalization of lattice gauge Hamiltonians is constrained by the
requirements of gauge invariance. \ To preserve gauge invariance it is most
convenient to use a basis which diagonalizes the electric field part of the
Hamiltonian%
\begin{equation}
\left[  -\sum_{l}\tfrac{\partial^{2}}{\partial A_{l}^{2}}\right]
\bigotimes_{l^{\prime}}\left|  n_{l^{\prime}}\right\rangle \cdots=\left[
\sum_{l}n_{l}^{2}\right]  \bigotimes_{l^{\prime}}\left|  n_{l^{\prime}%
}\right\rangle .
\end{equation}
As our next example of stochastic error correction we will address the
$4\times4$ lattice system at $x=1$ using this electric field basis. \ In
\cite{irving} it was noted that this poses a challenge to standard
diagonalization techniques. \ Even on the small $4\times4$ lattice a
surprisingly large number of states, about $10^{7}\sim10^{8}$, are needed to
accurately describe the low energy spectrum at $x=1$. \ This problem can be
circumvented by modifying the basis states to incorporate more of the physics
of the ground state. \ For example one can introduce a disordered background
of magnetic flux as suggested in \cite{heys}, and that approach is followed in
an ongoing project \cite{gauge}. \ However we would like to directly address
the problem described in \cite{irving} and show how the stochastic Lanczos
method handles the proliferation of large numbers of basis states in the
original electric field basis.

We will choose our starting subspace $S$ to include all basis states
\begin{equation}
\bigotimes_{l^{\prime}}\left|  n_{l^{\prime}}\right\rangle
\end{equation}
which satisfy%
\begin{equation}
\sum_{l}n_{l}^{2}\leq8,
\end{equation}
and which can be reached from the strong coupling vacuum by at most two
transitions via the plaquette operators $\exp(\pm i\theta_{P})$. \ We take the
auxiliary space $Z$ to be the subspace spanned by basis vectors%
\begin{equation}
\sum_{l}n_{l}^{2}\leq L_{\max}^{2}.
\end{equation}
Using matrix diffusion Monte Carlo, we diagonalize the subspace formed by the
states%
\begin{equation}
\left|  d,j\right\rangle =\left[  P_{Z}(H-a)P_{Z}\right]  ^{d}\left|
j\right\rangle .
\end{equation}
$\left|  j\right\rangle $ are the eigenvectors of the Hamiltonian restricted
to the original subspace $S$. \ In our calculations we use $a=L_{\max}^{2}$
and $d=0,1,\cdots12$ for cutoff values,
\begin{equation}
L_{\max}^{2}=24,28,32.
\end{equation}
In Table 2 we show the results for the ground state energy $E_{0}$ for the
different cutoff values $L_{\max}^{2}$ and the extrapolated value at $L_{\max
}^{2}=\infty$. \ The errors shown are estimated statistical errors. \ For
comparison we show the results of \cite{hamer} obtained using Green's function
Monte Carlo (GFMC).%

\[
\overset{\text{Table 2}}{%
\begin{tabular}
[c]{|l|l|l|l|l|l|}\hline
& $L_{\max}^{2}=24$ & $L_{\max}^{2}=28$ & $L_{\max}^{2}=32$ & $L_{\max}%
^{2}=\infty$ & GFMC\\\hline
$E_{0}$ & $-7.394(3)$ & $-7.430(3)$ & $-7.438(3)$ & $-7.442(4)$ &
$-7.4432(5)$\\\hline
\end{tabular}
}%
\]
In Table 3 we show the masses for the lightest six particles in the system
extrapolated to the limit $L_{\max}^{2}=\infty$. \ We have labelled the
particles according to their spin $J$ and sign under conjugation
$C:A\rightarrow-A.$ We also include results from \cite{hamer} for the lowest
antisymmetric and symmetric glueballs.%
\[
\overset{\text{Table 3}}{%
\begin{tabular}
[c]{|l|l|l|}\hline
$J^{C}$ & Mass & GFMC\\\hline
$\left|  0^{-}\right\rangle $ & $3.03(2)$ & $3.01(6)$\\\hline
$\left|  0^{+}\right\rangle $ & $4.03(3)$ & $4.05(8)$\\\hline
$\left|  2^{-}\right\rangle $ & $6.8(1)$ & \\\hline
$\left|  2^{+}\right\rangle $ & $6.8(1)$ & \\\hline
$\left|  0^{+}\right\rangle $ & $7.0(2)$ & \\\hline
$\left|  0^{-}\right\rangle $ & $7.1(2)$ & \\\hline
\end{tabular}
}%
\]
The results we find appear in agreement with \cite{hamer}. \ Unlike most Monte
Carlo algorithms, the SEC method is able to find excited states with the same
quantum numbers as lower lying states. \ This was also evident in the
$\phi_{2+1}^{4}$ example where we could track many different states crossing
the phase transition. \ The reason for this advantage goes back to the design
of stochastic error correction as a Monte Carlo improvement of a
diagonalization scheme. \ For the $U(1)$ example one can reliably find the
eigenvalues and eigenvectors for the first twenty or so states in the low
energy spectrum.

\section{Hubbard Model}

The last example we consider is the two-dimensional Hubbard model defined by
the Hamiltonian
\begin{equation}
H=-t\sum_{<i,j>;\;\sigma=\uparrow,\downarrow}(c_{i\sigma}^{\dagger}c_{j\sigma
}+c_{j\sigma}^{\dagger}c_{i\sigma})+U\sum_{i}(c_{i\uparrow}^{\dagger
}c_{i\uparrow}c_{i\downarrow}^{\dagger}c_{i\downarrow}).
\end{equation}
The summation $<i,j>$ is over nearest neighbor pairs. $\ c_{i\sigma}^{\dagger
}$($c_{i\sigma}$) is the creation(annihilation) operator for a spin $\sigma$
electron at site $i$. $\ t$ is the hopping parameter, and $U$ controls the
on-site Coulomb repulsion. The model has attracted considerable attention in
recent years due to its possible connection to $d$-wave pairing and stripe
correlations in high-$T_{c}$ cuprate superconductors. \ In spite of its simple
form, the computational difficulties associated with finding the ground state
of the model are substantial even for small systems. \ Fermion sign problems
render Monte Carlo simulations ineffective for $U$ positive and away from
half-filling, and the collective effect of very large numbers of basis Fock
states make most diagonalization approaches very difficult. A brief overview
of the history and literature pertaining to numerical aspects of the Hubbard
model can be found in \cite{overview}.

In terms of momentum space variables, the Hubbard Hamiltonian on an $N\times
N$ periodic lattice has the form%
\begin{align}
H  &  =-2t\sum_{p_{x},p_{y}=0,\cdots,N-1}(\cos\tfrac{2\pi p_{x}}{N}+\cos
\tfrac{2\pi p_{y}}{N})\left[  c_{p_{x},p_{y}}^{\uparrow\dagger}c_{p_{x},p_{y}%
}^{\uparrow}+c_{p_{x},p_{y},\sigma}^{\downarrow\dagger}c_{p_{x},p_{y}%
}^{\downarrow}\right] \\
&  +\tfrac{U}{N^{2}}\sum_{\substack{p_{x}-q_{x}+r_{x}-s_{x}%
=0\operatorname{mod}N\\p_{y}-q_{y}+r_{y}-s_{y}=0\operatorname{mod}N}%
}c_{p_{x},p_{y}}^{\uparrow\dagger}c_{q_{x,}q_{y}}^{\uparrow}c_{r_{x},r_{y}%
}^{\downarrow\dagger}c_{s_{x},s_{y}}^{\downarrow}.\nonumber
\end{align}
As a test of our methods, we use QSE diagonalization with stochastic error
correction to find the ground state energy of the $4\times4$ Hubbard model
with $5$ electrons per spin. \ The corresponding Hilbert space has about
$2\cdot10^{7}$ dimensions. \ For the QSE diagonalization we use momentum Fock
states which diagonalize the quadratic part of the Hamiltonian. \ The
Hamiltonian is invariant under the symmetry group generated by reflections
about the $x$ and $y$ axes, interchanges between $x$ and $y,$ and interchanges
between $\downarrow$ and $\uparrow$. \ We find it convenient to work with
symmetrized Fock states. \ We will compute stochastic error corrections to
first order using the series method.

In Table 4 we present results for the ground state energy. \ We encountered no
trouble with the sign problem, and in fact one can easily see that each term
in the first order series expression (\ref{firs}) is negative definite. \ The
energies are measured relative to the energy of the Fermi sea at $U=0$. \ The
errors reported are statistical errors associated with the first order SEC
calculation. \ Where available, we compare with the results presented in
\cite{huss}, which we label as Exact, Projector Quantum Monte-Carlo (PQMC),
and Stochastic Diagonalization (SD). \ Stochastic diagonalization is a
subspace diagonalization technique similar to QSE but one which uses a
different method for selecting the subspace and is based on a variational
principle \cite{deraedt}. \ Although the precise number of basis states used
in the SD calculations is not listed, we infer from numbers reported for a
modified $4\times4$ Hubbard system that roughly $10^{5}$ states were
used.\footnote{The discrete symmetries of the system were not utilized in
their calculations.}%

\[
\overset{\text{Table 4}}{%
\begin{tabular}
[c]{|l|l|l|l|l|l|l|}\hline
Coupling & States & QSE & QSE+SEC & Exact & SD & PQMC\\\hline
$U=2t$ & $%
\begin{array}
[c]{c}%
100\\
500\\
1000
\end{array}
$ & $%
\begin{array}
[c]{c}%
-.4797\\
-.4945\\
-.5006
\end{array}
$ & $%
\begin{array}
[c]{c}%
-.50147(5)\\
-.50181(3)\\
-.50198(1)
\end{array}
$ & $-.50194$ & $-.5010$ & $-.44(5)$\\\hline
$U=4t$ & $%
\begin{array}
[c]{c}%
100\\
500\\
1000
\end{array}
$ & $%
\begin{array}
[c]{c}%
-1.620\\
-1.748\\
-1.800
\end{array}
$ & $%
\begin{array}
[c]{c}%
-1.8113(4)\\
-1.8242(3)\\
-1.8302(1)
\end{array}
$ & $-1.8309$ & $-1.829$ & $-1.8(2)$\\\hline
$U=5t$ & $%
\begin{array}
[c]{c}%
500\\
1000\\
2000
\end{array}
$ & $%
\begin{array}
[c]{c}%
-2.558\\
-2.651\\
-2.685
\end{array}
$ & $%
\begin{array}
[c]{c}%
-2.7073(4)\\
-2.7208(2)\\
-2.7231(1)
\end{array}
$ & $-2.7245$ & $-2.723$ & $-2.9(3)$\\\hline
\end{tabular}
}%
\]
Apparently QSE diagonalization with SEC handles the $4\times4$ system quite
well with relatively few states. \ Much larger systems are being studied using
both higher series corrections and stochastic Lanczos techniques \cite{salwen}.

\section{Summary and comments}

In this paper we presented two versions of stochastic error correction, the
series method and the stochastic Lanczos method. \ The series method starts
with eigenvectors of the Hamiltonian restricted to some optimized subspace and
includes the contribution of the remaining basis states as an ordered
expansion. \ The stochastic Lanczos method starts with eigenvectors of a
Hamiltonian submatrix and constructs matrix elements of Krylov vectors using
matrix diffusion Monte Carlo. \ This method has the advantage that the
starting vectors need not be close to exact eigenvectors.

We presented three different examples which demonstrate the potential of the
new approach for strongly coupled scalar, gauge, and fermionic theories. \ In
the first example we calculated the low energy spectrum of $\phi_{2+1}^{4}$
using the series method, and in the second example we found the spectrum of
compact $U(1)$ in $2+1$ dimensions using the stochastic Lanczos method. \ In
both examples we found agreement with results from the literature. \ We also
found that unlike typical Monte Carlo results, the SEC method is able to find
the eigenvalues and eigenvectors for excited states with the same quantum
numbers as lower lying states. \ This advantage is due to its design as a
Monte Carlo improvement of a diagonalization scheme. \ In the last example we
found the ground state of the $2+1$ dimensional Hubbard model using QSE
diagonalization and first order series stochastic error correction. \ In this
calculation we encountered no fermion sign problem and found that our methods
yielded very accurate results with far less effort than existing techniques.
We believe that the methods we have presented hold considerable potential for
studying a wide range of non-perturbative quantum systems and answering
questions difficult to address using other methods.

\paragraph*{Acknowledgments}

We are grateful to Eugene Golowich, Daniel Lee, and Nikolay Prokofiev for
conversations. \ Financial support provided by the National Science Foundation.

\begin{figure}[t]
\begin{center}
\epsfxsize=25pc \epsfbox{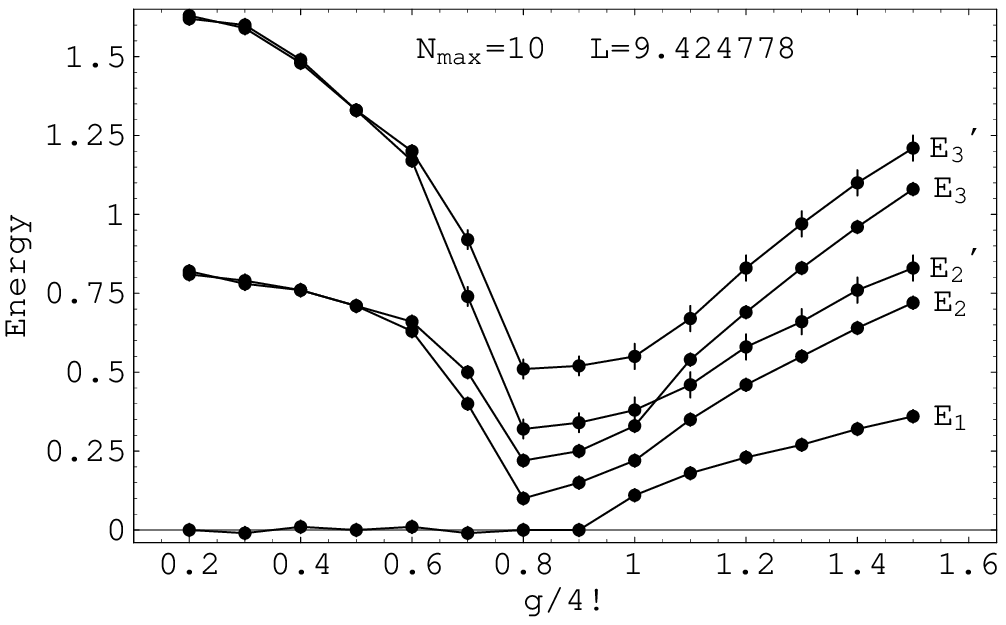}
\end{center}
\caption{Energy eigenvalues as functions of $\frac{g}{4!}$ as calculated by
QSE diagonalization with first-order error corrections.}%
\end{figure}
\end{document}